# Tuning the electron-phonon interaction via exploring the interrelation between Urbach energy and Fano-type asymmetric Raman line shape in GO-hBN nanocomposites

*Vidyotma Yadav, Tanuja Mohanty\**

*School of Physical Sciences, Jawaharlal Nehru University, New Delhi 110067*

**Abstract**

Hexagonal boron nitride (hBN), having an in-plane hexagonal structure in the $sp^2$ arrangement of atoms, proclaims structural similarity with graphene with only a small lattice mismatch. Despite having nearly identical atomic arrangements and exhibiting almost identical properties, the electronic structures of the two materials are fundamentally different. Considering the aforementioned context, a new hybrid material with enhanced properties can be evolved combining both materials. This experiment involves liquid phase exfoliation of hBN and two-dimensional nanocomposites of GO-hBN with varying hBN and graphene oxide (GO) ratios. The optical and vibrational studies conducted using UV-Vis absorption and Raman spectroscopic analysis report the tuning of electron-phonon interaction (EPI) in the GO-hBN nanocomposite as a function of GO content (%). This interaction depends on disorder-induced electronic and vibrational modifications addressed by Urbach energy ($E_u$) and asymmetry parameter ($q$), respectively. The EPI contribution to the induced disorders estimated from UV-Vis absorption spectra is represented as EPI strength ($E_{e-p}$) and its impact observed in Raman phonon modes is quantified as an asymmetry parameter ($q$). The inverse of the asymmetry parameter is related to $E_{e-p}$, as $E_{e-p} \sim 1/|q|$. Here in this article, a linear relationship has been established between $E_u$ and the proportional parameter (k), where k is determined as the ratio of the intensity of specific Raman mode ($I$) and $q^2$, explaining the disorders' effect on Raman line shape. Thus a correlation between Urbach energy and the asymmetry parameter of Raman mode confirms the tuning of EPI with GO content (%) in GO-hBN nanocomposite.

# 1. Introduction

hBN is a unique two-dimensional (2D) material possessing high chemical and thermal stability, having a large band gap (~ 6 eV) and large thermal conductivity, thus providing access to a wide range of potential applications in optoelectronics, deep ultraviolet detector, photoconductivity, photocatalysis, photodetectors and device design etc. [1-4]. Understanding a material's optical and electronic characteristics and studying structural defects is necessary to enable such applications. In hBN, boron and nitrogen atoms are assembled in a hexagonal lattice at adjacent sites in the sp$^2$ configuration [5]. The fact that hBN is electrically insulating due to its high band gap, which was once considered a drawback from the application perspective, has now become a strength in the field of device fabrication. hBN's structural similarity with graphene and graphene-based materials such as GO provides suitable tuning of the electronic and optical properties of hBN [6-7]. It is possible to evolve a new hybrid material with improved qualities from these two materials with similar structures but different electronic properties. The knowledge and management of h-BN's electrical and optical characteristics are crucial for its effective use, much like with other semiconductors of technical significance. This work uses an efficient technique to synthesize nanocomposites of hBN and GO with five different ratios. The crystal structure of hBN is expected to be modified chemically due to the introduction of the GO domain and synthesis by-products [7]. The hBN and graphene oxide hybrid is expected to produce semiconductors with variable band gaps for different ratios. Due to compositional fluctuation in the crystal lattice, the chemical and structural influence on the lattice can induce disorders such as vacancies, edge defects, adatoms, edge irregularities, substitution, and inhomogeneity [1,8-10]. These disorders can affect the phononic as well as electronic structure of the material [8]. Disorder-induced electronic and phononic modifications can be addressed by the Urbach energy ($E_u$) and asymmetry parameter ($q$), respectively, that quantifies disorders present in the system [8, 11-12]. For that purpose, a 2D nanocomposite material GO-hBN is chosen to observe the effect of electron-phonon interaction (EPI) with increasing GO content (%).

EPI is one of the elementary interactions of quasiparticles which influences phonon-assisted optical absorption in indirect bandgap semiconductors [13,14]. While significant work has been done in demonstrating the role of phonon vibrations on electron behaviour, a few observations have also been made to study the influence of electrons on phonon dynamics [13]. Tuning the response of both phonons and electrons can effectively modify the intrinsic optical properties of a 2D material, which assists in enhancing the performance of optoelectronic devices [15]. The EPI significantly impacts the light-matter interaction in 2D layered materials, Van der Waals heterostructures, etc. In layered semiconducting material, the electrons and phonons can interact within the layer (intralayer EPI), or there is a possibility of phonons interacting with electrons in an adjacent layer (interlayer EPI) [16]. In a layered nanocomposite, a systematic variation in the EPI is expected with varying ratios of components [17]. It has been shown that the variation of components affected the disorder-induced behaviour of EPI. The symmetry of the electronic and phononic states at the minima of the conduction band (CBM) and maxima of the valence band (VBM) is essential for the phonon renormalization caused by a variation of components [17]. The EPI is a vital scattering phenomenon that affects electron and phonon transport which can be studied by employing UV Visible absorption and Raman spectroscopy, respectively [18]. UV-Vis absorption spectroscopy examines the material's electronic properties, such as Optical bandgap, Urbach energy, etc. EPI is expected to affect a disorders-related parameter, Urbach energy which can be estimated through the optical absorption coefficient following the relation. $\alpha = \alpha_0 exp\left[\frac{h\nu}{E_u}\right]$ [19].

The optical absorption coefficient gets affected for 2D materials where the quantum confinement effect plays a significant role. It generally has a higher value than its bulk crystalline form [19]. In addition to the quantum confinement effect, the optical absorption coefficient's exponential response to incident photon energy near absorption edges is believed to be originated due to EPI, which will have a consequent impact on Urbach energy. The vibrational properties of semiconductors are studied by analyzing the FWHM, shift in peak position, and intensity of a specific mode of Raman spectra. In Raman spectroscopy, the material undergoes an inelastic scattering while interacting with incident photons. In crystalline bulk semiconductors, only the phonons at the zone centre (k = 0) participate in Raman scattering, conserving its momentum to give rise to a symmetric Raman profile. However, there is a relaxation in the Raman selection rule in a low-dimensional semiconducting material due to quantum confinement. As a result, not only phonons at the zone centre but also the other phonons near the zone centre (k > 0) contribute to scattering, thus leading to an asymmetric broadening of the Raman profile [18,20].

Along with the quantum confinement effect, one more phenomenon called the Fano resonance effect also impacts the symmetry of the Raman profile. The Fano effect results in a characteristic asymmetry in the Raman profile of the semiconductors [21]. The interaction strength of electrons with lattice vibrations is studied via the asymmetric profile of Raman mode. The EPI is determined in terms of Raman parameters such as frequency of Raman mode ($\omega$), linewidth parameter ($\Gamma$), and intensity of specific Raman mode (I($\omega$)) [17]. The EPI strength has been tuned as a function of GO content (%). The asymmetry in the Raman line shape arises when the defect-generated discrete states interact with a continuum of states, as depicted in Figure. 2(a)-(c). This type of interaction between discrete and continuum states is referred to the Fano resonance. The tuneable Fano resonance provides a new path for the sensor industry, optoelectronics and other device applications [17]. Numerous experimental and theoretical research has been done on the Fano profile of various materials, but no similar report has been observed on 2D nanocomposites of GO-hBN. In the present work, we have tried to tune the EPI systematically with variations of GO content (%). Therefore, our work has focused on the behaviour of EPI in terms of asymmetry parameter and Urbach energy.

Urbach energy can be evaluated by analyzing the UV-Visible absorption spectra of nanocomposites. The characteristic Urbach energy can be expressed through three different contributions from structural, chemical, and thermal disorders. It can be presented as $E_u = E_X + E_C + E_T$ [8,11]. $E_X$ and $E_C$ are temperature-independent static structural and compositional disorders, respectively. $E_T$ is a temperature-dependent parameter that contributes to Urbach energy and is termed as a thermally induced disorder. These disorders can be generated in the system either during synthesis or through post-synthesis procedures. An increase in defects/disorders causes the energy levels to extend from the band edge into the forbidden energy gap, widening the band tail as depicted in Figure 1. The larger $E_u$ values result from the expansion of the Urbach band tail as a result of increase in compositional disorders. For crystalline materials, the Urbach law exhibits an exponential dependence of the absorption coefficient ($\alpha$) on photon energy and is given by: $\alpha = \alpha_0 exp\left[\frac{h\nu}{E_u}\right]$ [19,22-25].

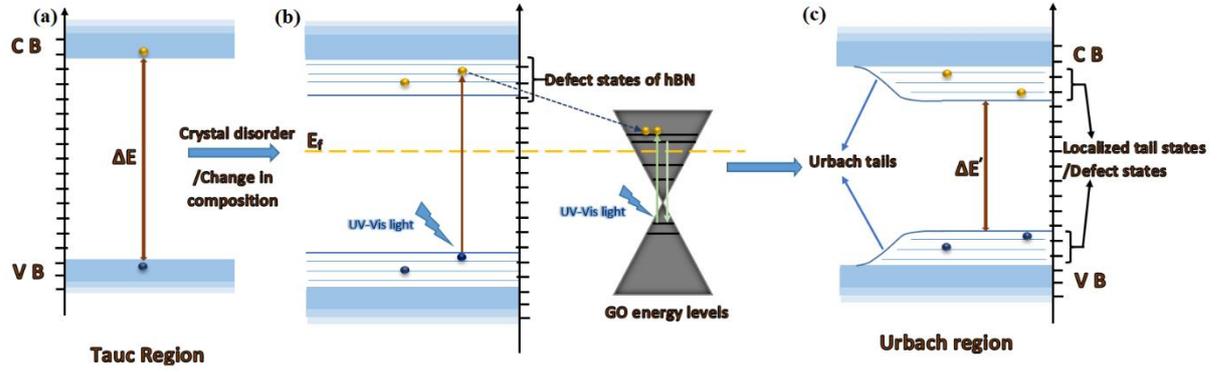

**Figure 1.** Schematic of (a) energy band diagram of hBN (b) energy band diagram of GO-hBN nanocomposite and electron transfer pathways (c) Generation of Urbach tails and band gap variation due to defect introduction.

Simultaneously, we expect that induced disorders' impact on phononic interactions alters nanocomposites' vibrational properties. Although no direct effect is observed between the induced disorders and electron-phonon interaction, the chemical and structural disorders impact both the phononic and electronic energy band states [26]. The system's electron-lattice interaction is also predicted to be influenced by the phonon spread [26]. The full-width half maxima (FWHM) of the phonon mode provides information about the phonon spread, which is already integrated into the Fano asymmetric parameter $q$ [26-28]. The Fano asymmetry parameter is a crucial parameter to get insight into disorders induced in the system through asymmetry present in a specific Raman mode. The asymmetric line shape in Raman spectra is emerged due to interaction between discrete bands and continuum bands, as depicted in Figure 2. The asymmetric line shape of the Raman mode incorporates a peak associated with resonance maxima and a dip/minimum called an anti-resonance dip [29]. The asymmetric line shape is analyzed to observe how resonance affects the electrical and vibrational properties. The asymmetric Raman bands are originated due to the Fano resonance scattering between phonons and conduction band electrons and can be represented by the Fano line-shape function: [27, 30].

$$I(\omega) = A \frac{[q + (\omega - \omega_o)/\Gamma]^2}{1 + [(\omega - \omega_o)/\Gamma]^2}$$

Where $A$ is the scaling factor, $q$ is the asymmetry parameter, $\omega_o$ is the experimentally observed Raman mode frequency in the presence of the coupled scattering, and $\Gamma$ is the experimental line width of the Raman mode, which is related to the phonon lifetime. The parameter $[\xi(\omega) = (\omega - \omega_o)/\Gamma]$ represents the reduced energy variable [31]. The value of $q$ determines how the electronic Raman scattering affects the phonon scattering. A lower value of $q$ indicates a stronger interference. The inverse of the asymmetry parameter provides the electron-phonon interaction strength, i.e., $E_{e-p} \sim \frac{1}{|q|}$ [30,32]. In contrast to the symmetric Lorentz profile, the symmetry of the Fano line shape depends on the value of the $q$ parameter. The discrete signal intensity will have a higher spectral dominance for higher $q$ values. In the absence of coupling between the external perturbation and the continuum state (i.e. When $q$ has such a high value that $q \to \pm\infty\ or\ \frac{1}{|q|} \to 0$ ), the Fano resonance line-shape transforms into a symmetric Lorentz line-shape function with $I(\omega) \propto \frac{1}{1+\xi^2}$ [30, 33-34] as shown in Figure 2(d). There is one more significant case where the Fano profile transforms into a symmetric quasi-Lorentz anti-resonance/inverted Lorentzian line-shape profile for $q = 0$, as shown in Figure 2(d), which indicates that the discrete state and external perturbation are not coupled. The inverted Lorentzian function can be represented as $I(\omega) \propto \frac{|\xi|^2}{1+\xi^2}$ [33].

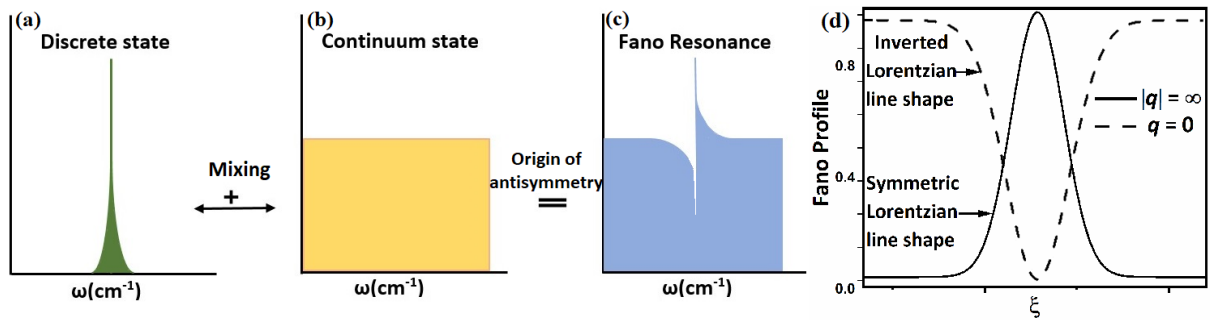

**Figure 2.** Schematic diagram of the emergence of the Fano resonance effect with the representation of (a) discrete phononic state, (b) continuum of electronic states, (c) antisymmetric mix states, (d) special cases of Fano line shape for $q = 0$ and $q = \infty$.

Here, in this work, both the UV-Vis absorption and Raman spectroscopic measurements on the nanocomposites have been carried out to give insight into the EPI. The $E_u$ parameter, estimated by analyzing UV-Vis spectra, presents a quantitative approach to studying the induced disorders in the samples. In contrast, the Fano asymmetry parameter ($q$) estimated from Raman spectra presents a qualitative approach to exploring the EPI [27]. The behaviour of both parameters $q$ and $E_u$ is supposed to have a common origin, i.e., due to EPI. The asymmetry parameter is also related to the spectral width of the interaction band ($L$) [27, 35]. In addition, the spectral width of the specific Raman mode can be expressed in terms of $E_u$, i.e. $L \propto E_u$ [36, 37]. Even though few reports have been observed on the optical properties of GO-hBN nanocomposites, no report has correlated its Urbach energy with the Fano asymmetry parameter. This article reports a linear relation between $E_u$ and a proportional parameter ($k$), where parameter $k$ is defined as the ratio of specific Raman mode intensity ($I$) and $|q|^2$ as $k = \frac{I}{|q|^2}$. Moreover, the correlation between $E_u$ and $k$ can serve as a diagnostic tool to monitor changes in a material's structural disorder and defects over time. This relation is beneficial for studying the physical phenomenon behind EPI and modifying $E_{e-p}$ through disorders [26].

## 2. Experimental details

### 2.1. Sample Synthesis

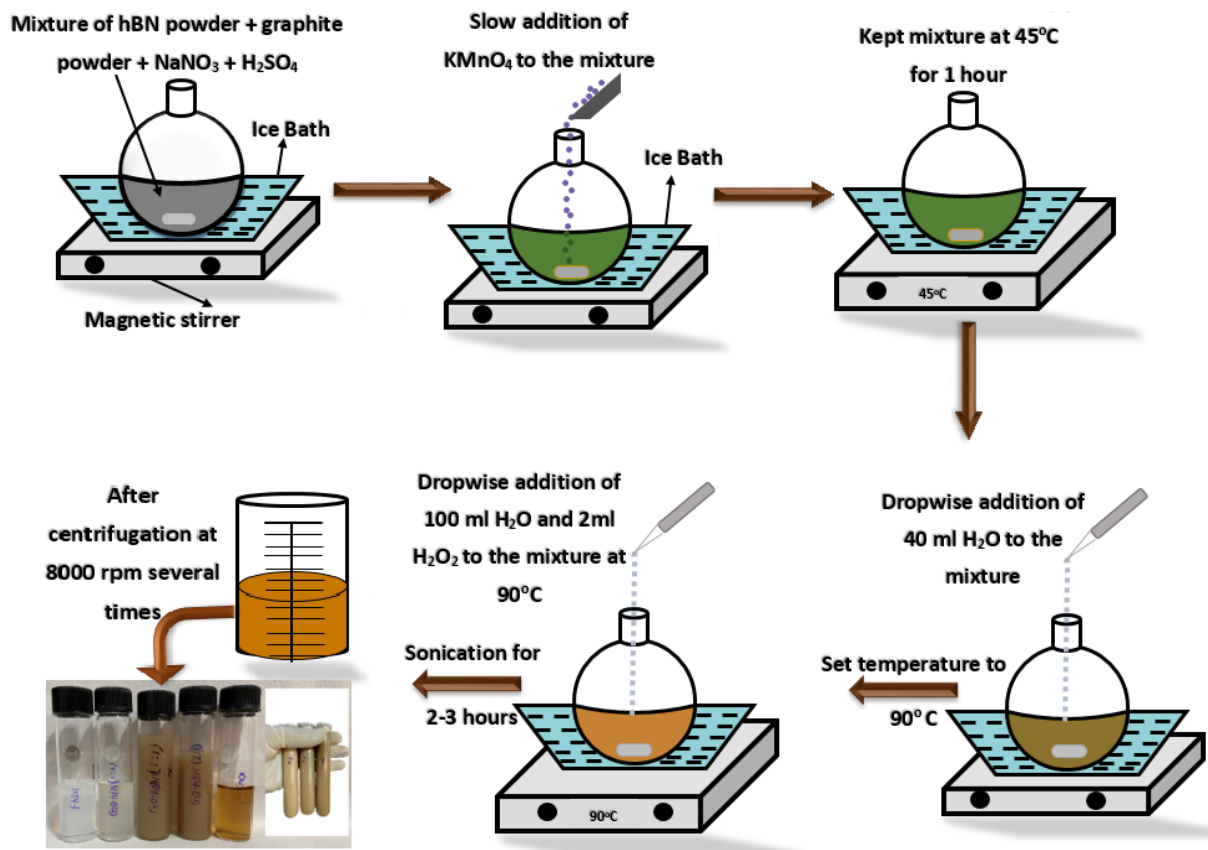

**Figure 3.** Schematic of stepwise synthesis procedure of GO-hBN nanocomposites.

This work presents the synthesis of GO-hBN nanocomposites using a modified Hummer method. Pristine hBN, GO, and GO-hBN nanocomposites with varying ratios of GO and hBN such as (1:3, 1:2, 1:1, 2:1 and 3:1) were synthesized [38]. For this purpose, graphite and hBN nanopowder were mixed in different ratios (0:1, 1:3, 1:2, 1:1, 2:1, 3:1, and1:0). The obtained solution was then sonicated for a few hours and stored for further purification. The solution was then centrifuged initially at 1000 rpm for 5-10 minutes to collect the supernatant and then multiple times at 8000 rpm for 30 minutes to collect the sediment in DI water. The final samples were deposited on Si substrate using the drop cast method for further investigation.

The schematic of GO-hBN nanocomposites is presented in Figure 4, where the GO domain is introduced into the hBN structure. In GO-hBN nanocomposite, the GO and hBN domains are not expected to connect via covalent bonds between carbon and nitrogen or carbon and boron atoms; instead, they are connected through functional groups.

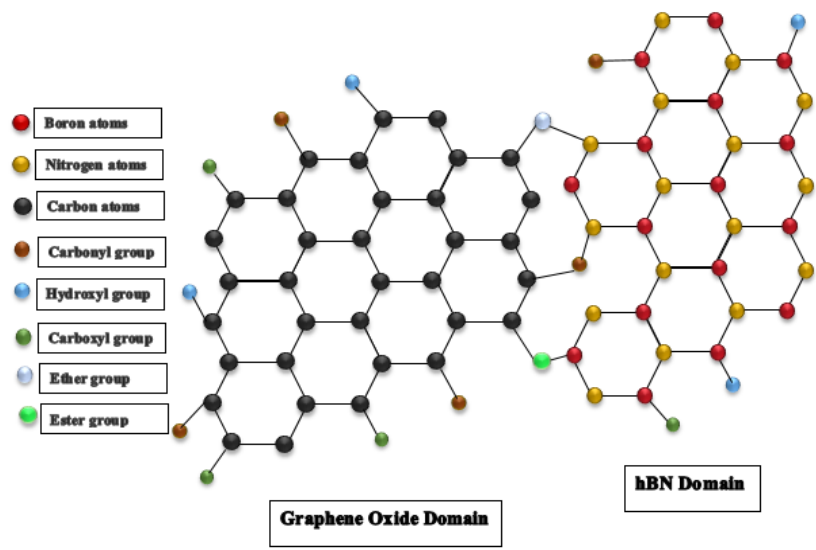

**Figure 4.** Schematic of GO-hBN nanocomposite.

**2.2. Characterization Techniques**

The synthesized GO-hBN nanocomposite samples were characterized by a scanning electron microscope (SEM, Zeiss EVO 40) to analyze the morphological features. The non-destructive technique Raman spectroscopy was carried out using a Wi-Tech alpha 300 confocal micro-Raman spectrometer with 532 nm laser excitation to investigate optical response and EPI on the samples. Optical bandgap modulation of samples and Urbach energy estimation were carried out with UV-Vis absorption spectroscopy using SHIMADZU UV-Vis 2600 spectrometer.

**3. Results and discussion**

**3.1. Scanning Electron Microscopy (SEM)**

SEM was used for the morphological study of the nanocomposites. The magnification was kept constant to collect the SEM images for GO and the nanocomposites; as shown in Figure 5, a different magnification was used to collect hBN's image for better visualization. The aggregated behaviour of nanoparticles of hBN is displayed in Figure 5(a). hBN has an impeccable nanosheet structure with a smooth surface, as clearly visible in Figure 5(a). GO exhibits a multilayer and fluffy structure, as shown in Figure 5(e). SEM images for the nanocomposites are shown in Figure 5(b) to 5(e), where the presence of both the multilayer structure of GO and nanosheets of hBN is observed. Also, as the GO content of the nanocomposite is increased, the increment in the GO multilayer structure is visible in the images. For the nanocomposite, GO-hBN (1:2), hBN nanosheets cover more surface than GO, as shown in Figure 5(b), whereas, in Figure 5(d) and 5(e), GO multilayer covers more surface in case of the nanocomposites with GO-hBN ratio as (2:1) and (3:1).

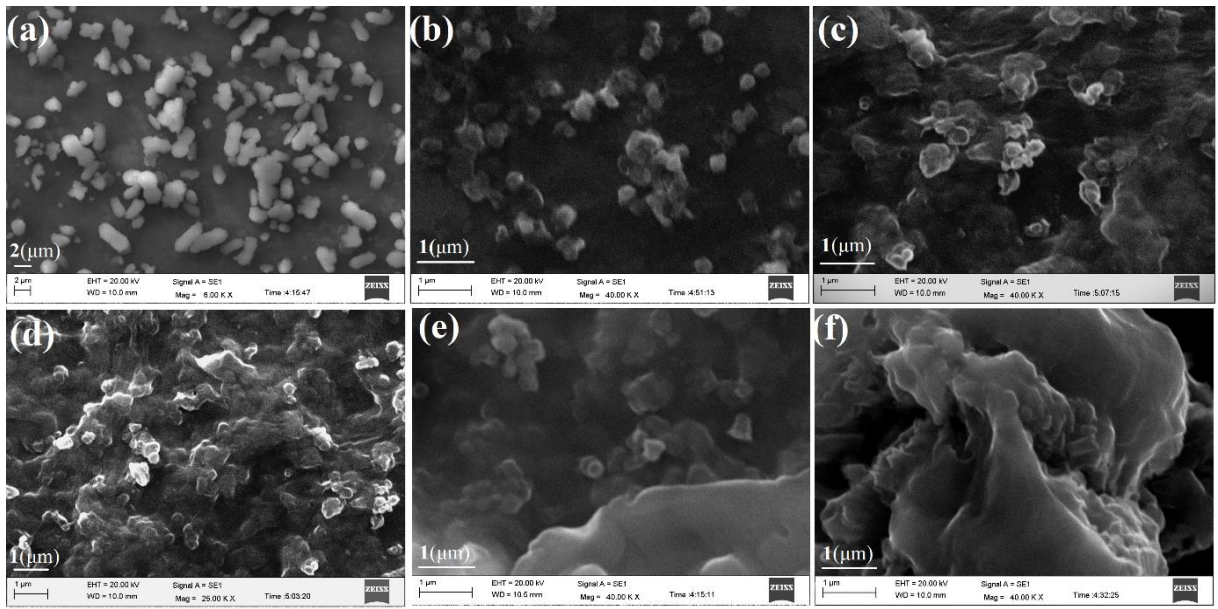

**Figure 5.** SEM images of (a) hBN (b) GO-hBN (1:2) (c) GO-hBN (1:1) (d) GO-hBN (2:1), (e) GO-hBN (3:1) and (f) GO.

## 3.2 UV-Visible Absorption Studies

UV-Vis absorption spectroscopy was performed on all samples in the wavelength range of 190 nm to 800 nm, and the spectra are presented in Figure 6(a). The sharp and intense absorption peak for hBN was noticed in the deep ultraviolet range, approximately at ~ 192 nm, resulting from $\pi$ to $\pi^*$ transition in hBN nanosheets [39]. The characteristic absorption peaks of GO were observed at 230 nm, indicating the $\pi$-$\pi^*$ transition between carbon atoms. Another peak at 305 nm suggests the n-$\pi^*$ transition between carbon and oxygen [4]. GO-hBN nanocomposites exhibited absorption peaks identical to the characteristic peaks of hBN and GO peaks, demonstrating the synthesis of GO-hBN nanocomposite materials. Out of two prominent peaks of GO in the absorption spectra for nanocomposites', one is in the range of ~ 230 nm - 250 nm, which corresponds to π to π* transition and another shoulder peak present in the range of ~ 280 nm - 320 nm corresponds to n to π* transition, indicating oxidation of graphene sheets [40]. Also, the signature peak of hBN can be seen to some extent in the nanocomposites. Additionally, the bandgap was estimated using Tauc's relation, as mentioned in equation (1) [20,25,41].

$$\alpha h\nu^n = A(h\nu - E_g) \qquad (1)$$

Where $\alpha$ is the absorption coefficient, $h$ is the Plank's constant, $h\nu$ is the photon energy of the source employed, $A$ is the proportionality constant, and $E_g$ is the optical energy bandgap of the material [42]. The n parameter relates to a specific electronic transition associated with light absorption. Its value is 2 or ½ depending on whether the material has a direct or indirect bandgap, respectively [42,43]. Considering the materials as indirect bandgap material [44,45], the calculated bandgaps for hBN, GO-hBN (1:3), GO-hBN (1:2), GO-hBN (1:1), GO-hBN(2:1), GO-hBN (3:1) and GO are found to be 5.05 eV, 2.50 eV, 2.44 eV, 2.10 eV, 2.30 eV, 2.54 eV and 3.11 eV, respectively as given in Table 1.

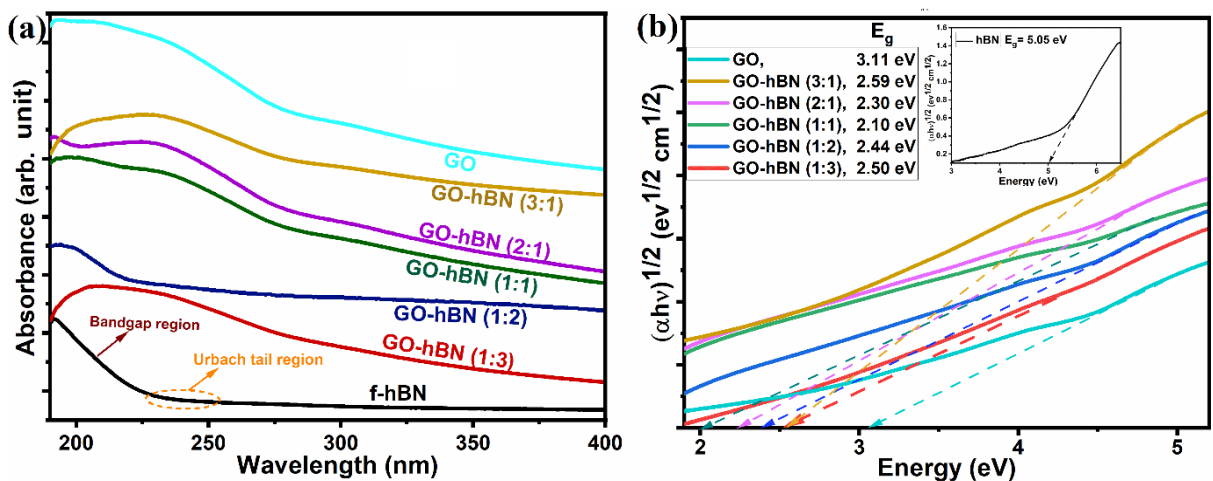

**Figure 6.** (a) UV-Vis absorption spectra of hBN and the GO-hBN nanocomposites (b) Optical bandgap of GO-hBN nanocomposites with GO content (%) using Tauc's relation (inset: Bandgap of hBN using Tauc's plot). Indirect bandgap was attributed to hBN and GO domains in the nanocomposites. The intercept of the dash lines with the horizontal axis elucidates the bandgap value.

The decrement in the bandgap of hBN with increasing GO content is supported by previously reported results [46]. The nanocomposites of hBN and GO are observed to have optical bandgaps that deviate from that of hBN. The bandgaps for nanocomposites decrease as the GO content increases from 0 to 50 %, and then it starts rising with the increasing GO content. The peculiar bandgap trend shown in Figure 7(b) may be explained through internal transitions within the bandgap as the GO content varies in the hBN lattice. The bandgap decrement is initially attributed to the insertion of GO domain in the nanocomposite resulting in the emergence of an intermediate structure between hBN and GO up to GO-hBN (1:1) ratio [6]. The minimum band gap is estimated for a particular ratio GO-hBN (1:1), where both GO and hBN are present almost in the same amount. As GO content starts increasing beyond 50% for samples GO-hBN (2:1 and 3:1), it is expected that the covalent functionalization of hBN with GO gets modified [47]. The modification in covalent functionalization of nanocomposites restricts the further reduction in bandgap with increasing GO content beyond 50%. As a result, the bandgap of nanocomposites gets modified in such a way that both the conduction and valence bands move upward, resulting in the net increase in the bandgap of GO-hBN nanocomposites (2:1 and 3:1) [48].

The band gap of material is also affected by the introduction of disorders. The study of disorders can be done by an energy parameter named Urbach energy through UV-Vis spectra of a material. Urbach energy can be extracted using the Urbach law mentioned in equation (2) and extrapolating the straight line of plot ln($\alpha$) versus $h\nu$. The inverse of the slope of the straight line will give us values of $E_u$ of pristine hBN and the nanocomposites. $E_u$ values for the nanocomposites are found to be more than pristine hBN. Figure 7(a) represents the observed trend between the estimated bandgap ($E_g$) and Urbach energy ($E_u$). From the evaluated value for both energy parameters, with a decrease in the bandgap, Urbach energy is found to be increased, as shown in Figure 7(a). A greater $E_u$ value indicates the increased structural disorder in nanocrystalline material. A relation given by Urbach that correlates the absorption coefficient to the energy bandgap is expressed as [19]:

$$\alpha = \beta \exp\left[\frac{\sigma(h\nu)}{K_B T}\right] \qquad (2)$$

Where $\alpha$ is the absorption coefficient, $\beta$ is an exponential coefficient and $\sigma$ is a temperature-independent constant known as the steepness parameter [19].

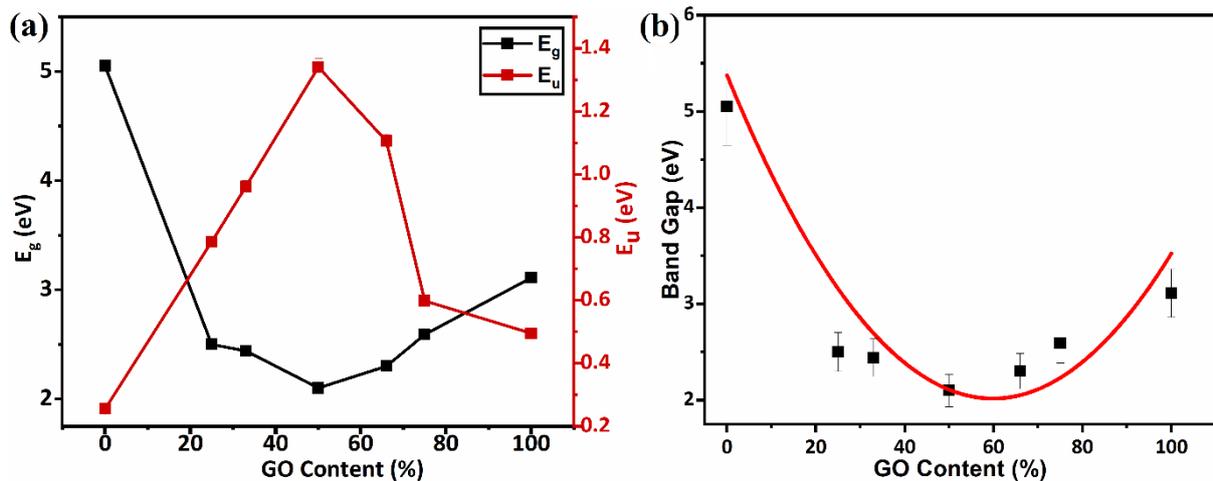

**Figure 7.** (a) Comparison of optical bandgap ($E_g$) and Urbach energy ($E_u$) of GO-hBN nanocomposites with GO content (%). (b) Curve fitting of bandgap in the nanocomposites with GO content (%).

From relations (1) and (2), the value of the steepness parameter can be evaluated to represent in the following form: $\sigma = K_B T/E_u$ [19,23]. The value of $\sigma$ is inversely related to a parameter named electron-phonon interaction strength ($E_{e\text{-}p}$) and is expressed as [23]:

$$E_{e-p} = \frac{2}{3\sigma} = \frac{2E_u}{3K_B T} \tag{3}$$

The EPI strength in a system is a measure of how strongly electrons interact with lattice vibrations. EPI is measurable in terms of the $E_u$. A high $E_u$ value represents strong electronic and phononic interaction.

Here, the Urbach energy of the nanocomposites is reported to have an unusual trend. The increment in Urbach energy of nanocomposites can be explained by the defects induced due to structural disorders in the hBN lattice. As a result of the induced disorders, some additional levels are introduced near the conduction band, valence band, or both [26]. The lattice distortion level, i.e. the defects in nanocomposites increase as GO content rises to 50%, which corresponds to the ratio GO-hBN (1:1). An increment in defects leads to an extension of energy levels from the band edge into the forbidden energy gap that broadens the band tail. The broadening in the Urbach band tail is responsible for the higher $E_u$ values [23,49]. Widening at the edge of the Urbach tail is attributed to an increment in compositional disorders [11]. The $E_u$ value starts decreasing for the nanocomposites as GO content increases beyond 50 %. The drop in $E_u$ is observed due to the presence of fewer disorders than GO-hBN (1:1).

## 3.3. Raman Studies

The Raman spectra presented in Figure 8 were measured to study the microstructural vibrational behaviour of hBN nanosheets, GO-hBN nanocomposites and GO through their EPI. The intense peak at ~ 1364 cm$^{-1}$ is the only characteristic peak of hBN, representing the in-plane stretching vibrations corresponding to the E$_{2g}$ mode of hBN [50]. The number of layers of synthesized hBN is estimated with the formula $<N> = \frac{17.2}{\Gamma - 8.5 - 1.19\,logP} - 1$, which comes out to be ~4-5 layers [44,51], thus confirming its indirect band gap nature. hBN's characteristic peak is located at a similar position as the D band of GO, but its behaviour is identical to the vibrations associated with the G band of GO [50,52]. The characteristic peak of hBN with a noticeable split was accompanied in the nanocomposites a peak that mimics the D peak of GO. The peak corresponding to the G band indicates the stretching vibrations of C-C sp$^2$ bonded graphitic structure, i.e. in-plane stretching vibrations of sp$^2$ carbon bonds near $\Gamma$ point of the Brillouin zone [53]. The G band is present at ~ 1608.40 cm$^{-1}$ and ~ 1596 - 1600 cm$^{-1}$ in GO and the nanocomposites, respectively [54-56].

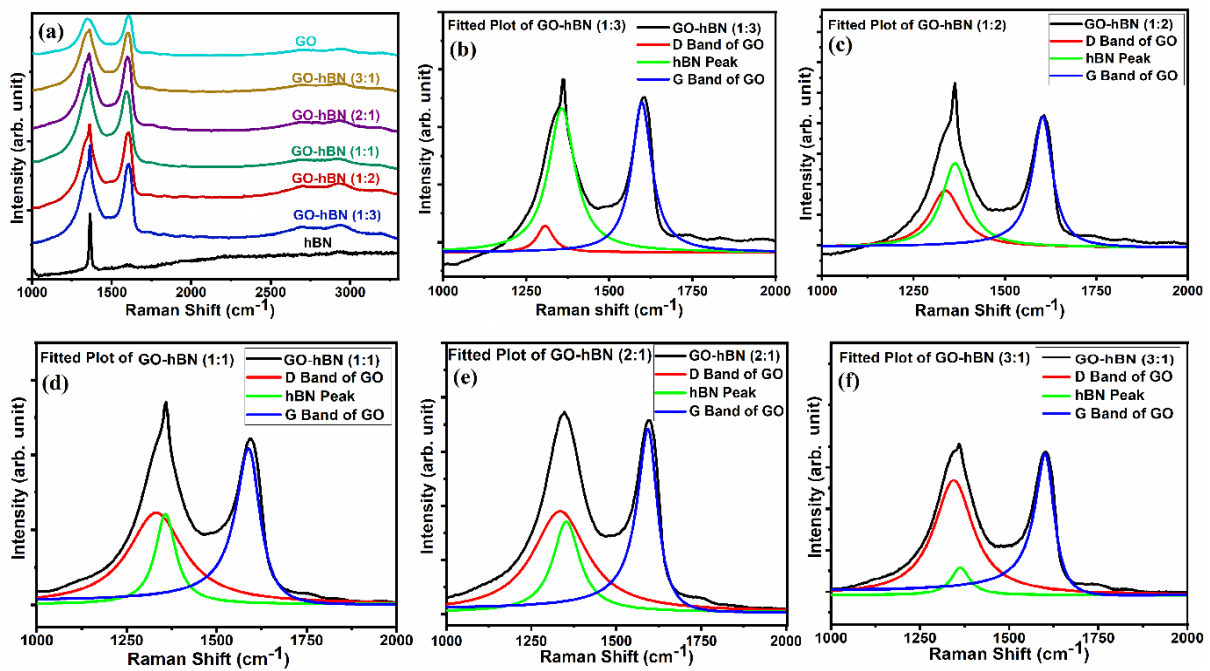

**Figure 8.** Raman spectra of (a) hBN, GO and the nanocomposites (b)-(f) BWF fitting of Raman spectra of the nanocomposites as (b) GO-hBN (1:3) (c) GO-hBN (1:2) (d) GO-hBN (1:1) (e) GO-hBN (2:1) (f) GO-hBN (3:1).

We have deconvoluted the recorded spectra for the nanocomposites into three different peaks. The fitted peak intensity of all three peaks gives the idea about the relative amount of hBN and GO content in the nanocomposite. The decrease in hBN content can be demonstrated by a progressive reduction in the intensity of the hBN peak with an increase in GO content in the Raman line shape fitted curves given in Figure 8(b) to 8(f).

In the Raman spectra (Figure 8), the experimental line shape of the $E_{2g}$ phonon mode is not symmetrical because of EPI [26]. A distinctive approach other than Lorentzian and Gaussian fitting was employed to consider this asymmetry. The Breit-Wigner-Fano resonance line shape (also called the Fano resonance or BWF resonance) fits the $E_{2g}$ phonon mode data, and this fitting is represented by the Fano function $I(\omega)$ [27,53]. We have evaluated $q$ values for the $E_{2g}$ phonon mode of hBN in GO-hBN nanocomposites by employing Fano fit to the spectra and observed its variations compared to pristine hBN. All the nanocomposites have a lower $q$ value than hBN, as presented in Table 1.

**Table 1.** Calculated values of $E_g$, $E_u$ and $q$ of pristine hBN and GO-hBN nanocomposites with different GO content (%).

| Sample | ($E_g$) eV | ($E_u$) eV | Asymmetry parameter $|q|$ |
|---|---|---|---|
| hBN | 5.05 | 0.2556 | 7.07 |
| GO-hBN(1:3) | 2.50 | 0.786 | 1.374 |
| GO-hBN(1:2) | 2.44 | 0.9615 | 1.8 |
| GO-hBN(1:1) | 2.10 | 1.34156 | 1.254 |
| GO-hBN(2:1) | 2.30 | 1.106 | 1.28 |
| GO-hBN(3:1) | 2.59 | 0.598 | 2.48 |
| GO | 3.11 | 0.49 | - |

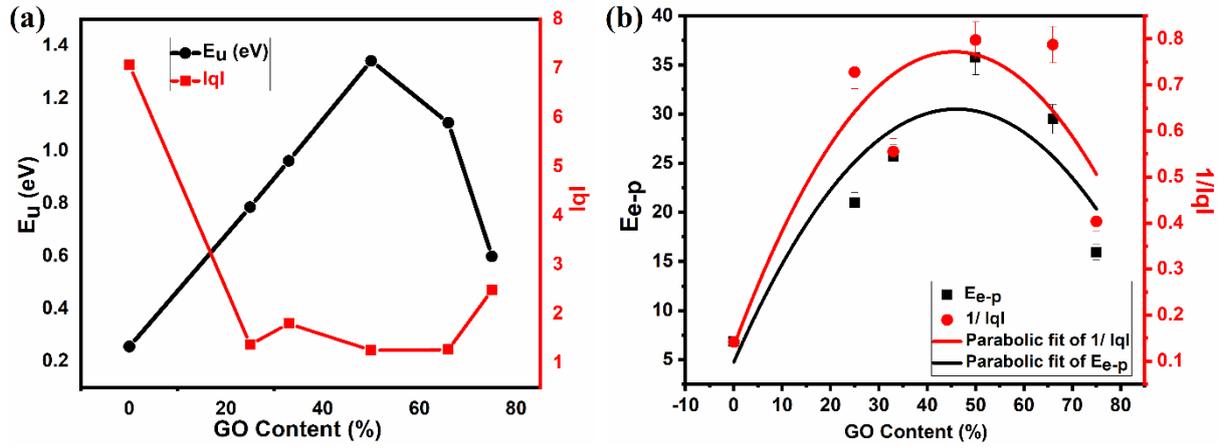

**Figure 9.** (a) Comparison of $q$ and $E_u$ with GO content (%). (b) The parabolic trend of $E_{e\text{-}p}$ in terms of $E_u$ (i.e. $\frac{2E_u}{3 K_B T}$) and the strength of EPI (i.e. $\frac{1}{|q|}$) with GO content (%).

The lower $q$ value indicates that electrons and phonons have a stronger interaction in the nanocomposites [30,32]. Figure 9(a) shows the inverse trend of $E_u$ and $q$ with GO content. The parameter $E_{e\text{-}p}$ is determined with the help of $E_u$ using equation (3). Here we have plotted the variation of $E_{e\text{-}p}$ with GO content (%), as shown in Figure 9(b). We have also shown the interpretation of $1/|q|$ with GO content, as depicted in Figure 9(b). Both $1/|q|$ and $E_{e\text{-}p}$ vary similarly, having a parabolic trend with an increase in GO content (%), thus proving that these parameters for the nanocomposites are associated with disorders present in the system.

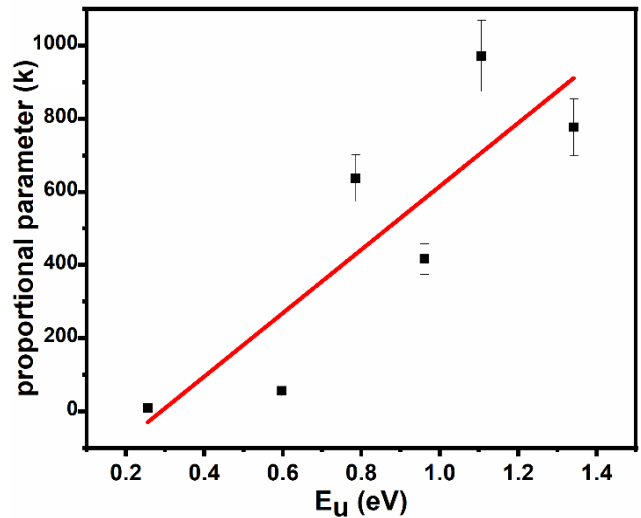

**Figure 10.** Linear relationship of $k$ parameter with $E_u$ in GO-hBN nanocomposites.

Using the calculated q values and intensity of Raman mode ($I$), we have determined a proportional parameter ($k$) defined as $k = I/q^2$ for all the nanocomposites. From the obtained values for $k$, we have plotted $k$ with Urbach energy, as shown in Figure 10. A linear relationship of $k$ with $E_u$ has been observed experimentally. This type of study has not been reported for GO-hBN before. This work reports a correlation between $E_u$ and $q$ for GO-hBN nanocomposites. Our experimental observed results are in good agreement with the theoretically estimated correlation as mentioned below in equations (15a) and (15b).

## 4. Theoretical model

In the present case, the hBN and its nanocomposites are considered to be in an atomic system having two different states: as a discrete phononic state ($|\Phi>$) and a continuum of electronic states ($|\Theta>$), as represented in Figure 2(a) and (b) respectively [32]. Here, $E_\Theta$ and $E_\Phi$ are the energy eigenvalues for the transitions in the state $|\Theta>$ and $|\Phi>$, respectively. Each state can be considered non-degenerate, where electrons and optical phonons interact with each other and can be represented in the following energy matrix form: [27,28].

$$<\Theta|\hat{H}|\Theta> = E_\Theta = \hbar\omega \quad \text{and} \quad <\Phi|\hat{H}|\Phi> = E_\Phi = \hbar\omega' \quad (4a)$$

Here, we examine the applicability of Fano's theory to the particular case. The fragments of the energy matrix are utilized to form another square submatrix which can be represented as equation (4b) after diagonalization. Considering the possibility of another continuum of energy states ($|\Theta>$) present in the system which is orthogonal to the energy states ($|\Phi>$) is presented as: [28]

$$<\Theta|\hat{H}|\Phi> = V_{E_\Theta} \quad (4b)$$

$$<\Theta'|\hat{H}|\Theta> = E_\Theta \delta(E_{\Theta'} - E_\Theta) \quad (4c)$$

Here, another state $|\Psi_\varepsilon>$ is defined as a mixed state, a combination of a discrete phononic state and a continuum of electronic states, as shown in Figure 2(c). The discrete phononic energy level $E_\Phi$ lies within the broad energy range of continuum levels $E_\Theta$. Also, each value ($\varepsilon$) is the eigenvalue of the matrix, which lies in the region of the continuum of energy levels $E_\Theta$. The eigenvector corresponding to that eigenvalue $\varepsilon$ can be expressed in terms of $|\Phi>$ and $|\Theta>$ as:

$$|\Psi_\varepsilon> = a|\Phi> + \int b_{E_\Theta} |\Theta> dE_\Theta \quad (5)$$

Where $a$ and $b_{E_\Theta}$ are energy ($\varepsilon$) dependent parameters and can be represented as [27]:

$$a = \frac{Sin\Omega}{\pi V_\varepsilon} \quad \text{and} \quad b_{E_\Theta} = \frac{V_{E_\Theta}}{\pi V_{E_\Theta}^*} \frac{Sin\Omega}{\varepsilon - E_\Theta} - Cos\Omega \, \delta(\varepsilon - E_\Theta) \quad (6)$$

Using these parameters in the mixed state $|\Psi_\varepsilon>$, equation (5) takes following form:

$$|\Psi_\varepsilon> = \frac{Sin\Omega}{\pi V_\varepsilon}|\Phi> + \int\left[\frac{V_{E_\Theta}}{\pi V_{E_\Theta}^*}\frac{Sin\Omega}{\varepsilon - E_\Theta} - Cos\Omega\, \delta(\varepsilon - E_\Theta)\right]|\Theta> dE_\Theta \quad (7)$$

where $\Omega$ represents the phase shift parameter due to the interaction between the states $|\Phi>$ and $|\Theta>$. An appropriate transition operator $\hat{A}$ is used to describe this transition of phonon from a state $|\kappa>$ within the energy range $E_\Theta$ to the mixed state $|\Psi_\varepsilon>$,

$$<\Psi\varepsilon|\hat{A}|\kappa> = \frac{Sin\Omega}{\pi V_\varepsilon^*}<P|\hat{A}|\kappa> - <\Theta|\hat{A}|\kappa> Cos\Omega \quad (8)$$

Where $/P>$ is a state defining the modifications in the phononic discrete state as: [28]

$$|P> = |\Phi> + \text{principle part of} \int \frac{V_{E_\Theta}|\Theta>}{\varepsilon - E_\Theta} dE_\Theta \quad (9)$$

Now to understand the intensity variation due to (i) electron-phonon interactions and (ii) disorders, the equation is rewritten as follows:

$$\frac{<\Psi_\varepsilon|\hat{A}|\kappa>}{<\Theta|\hat{A}|\kappa>} = \left(\frac{1}{\pi V_\varepsilon^*}\frac{<P|\hat{A}|\kappa>}{<\Theta|\hat{A}|\kappa>} - Cot\Omega\right) Sin\Omega \quad (10)$$

Considering the first term on the right side as asymmetry parameter ($q$) and the second term inside the parenthesis as reduced energy variable ($\xi$) of the equation (10) according to [27]. These terms are defined as:

$$q = \frac{1}{\pi V_\varepsilon^*}\frac{<P|\hat{A}|\kappa>}{<\Theta|\hat{A}|\kappa>} \quad (11)$$

$$\xi = -Cot\Omega = \frac{\varepsilon - E_\Phi - F(\varepsilon)}{\frac{1}{2}L} \quad (12)$$

where $L = 2\pi|V_\varepsilon|^2$ is the spectral width corresponding to how $|\Phi>$ interact with other states throughout a band of states, and $/V_\varepsilon/^2$ represents the configuration interaction. $F(\varepsilon)$ is the energy-dependent parameter representing the shift of resonance on either side w.r.t. the discrete energy level $E_\Phi$. Using values from equations (11) and (12) into equation (10), the ratio $\frac{|<\Psi_\varepsilon|\hat{A}|\kappa>|^2}{|<\Theta|\hat{A}|\kappa>|^2}$ is evaluated. Where $|<\Psi_\varepsilon|\hat{A}|\kappa>|^2$ is the transition probability to the mixed state and $|<\Theta|\hat{A}|\kappa>|^2$ is the transition probability to the continuum state.

$$\frac{|<\Psi_\varepsilon|\hat{A}|\kappa>|^2}{|<\Theta|\hat{A}|\kappa>|^2} = (q+\xi)^2 Sin\Omega^2 \quad or\quad = \frac{(q+\xi)^2}{1+\xi^2} \quad (13)$$

This ratio can be represented by a particular family of curves which are functions of the $q$ and $\xi$. Here, $q$ coincides with $\xi$ on either side of the resonance at a particular phase $\Omega = \Omega_o$, i.e., $E = E_o$, and the transition probability disappears on that side of the resonance.

Using the value as mentioned earlier of spectral width ($L$) and rewriting equation (13) as:

$$q^2 = \frac{1}{\pi^2 |V_\varepsilon|^2} \frac{|<P|\hat{A}|\kappa>|^2}{|<\Theta|\hat{A}|\kappa>|^2} \quad or \quad = \frac{2}{\pi} \frac{1}{L} \frac{|<P|\hat{A}|\kappa>|^2}{|<\Theta|\hat{A}|\kappa>|^2} \tag{14a}$$

The above relation can be rewritten in terms of the intensity parameter of Raman mode as:

$$\frac{q^2}{\frac{|<P|\hat{A}|\kappa>|^2}{|<\Theta|\hat{A}|\kappa>|^2}} = \frac{2}{\pi} \frac{1}{L} \tag{14b}$$

Where the intensity of a particular Raman mode ($I$) is correlated with the ratio of transition probabilities as $\frac{|<P|\hat{A}|\kappa>|^2}{|<\Theta|\hat{A}|\kappa>|^2} \sim I$. Notice that spectral width has dimensions of energy and remarkably depends on disorder width. Urbach energy is one possible outcome of these disorders. Thus, the spectral width can be expressed in terms of Urbach energy as: ($L \propto E_u$) [36, 37]. Defining a different parameter $k$ as the ratio of the intensity of Raman mode manifesting asymmetry ($I$) and $q^2$ as ($k = \frac{I}{q^2}$) and using this parameter in equation (14b),

$$\frac{1}{k} = \frac{q^2}{I} \propto \frac{1}{E_u} \tag{15a}$$

$$or \quad k \propto E_u \tag{15b}$$

This theoretically derived relation is proved experimentally in the present case, as shown in Figure 10. This relationship offers significant information regarding the degree of structural disorder and defects in a material.

## 5. Conclusions

In conclusion, the optical and vibrational studies of chemically synthesized pristine hBN and its nanocomposites with GO were carried out using UV-Vis absorption and Raman spectroscopy, respectively. Under specific circumstances, hBN nanosheets can be functionalized and transformed into highly conductive GO-hBN nanocomposite materials. Combining the benefits of GO and hBN can be intriguing, and the nanocomposite can act as a potential candidate for an enhanced electronic response, electrical and thermal conductivity.

We have observed the impact of chemically induced disorders on optical bandgap $E_g$ of hBN and GO-hBN nanocomposites. The dependence of both Urbach energy and asymmetry parameter of the nanocomposites on electron-phonon interaction is reported. The unusual behaviour of $E_g$ of nanocomposites with increasing GO content is expected due to the transitions in intermediate states as a result of modifications in covalent functionalization. The bandgap and Urbach energy are observed to have opposite trends with increasing GO content. A steep rise in $E_u$ values confirmed the increment of disorders in synthesized nanocomposites compared to hBN. The involvement of these disorders is reported to affect EPI and a parameter $E_{e-p}$ measures its strength. Also, the observed variation of EPI with disorders is explained in terms of asymmetry parameter $q$ of Raman modes. Thus the behaviour of $q$ and $E_u$ have a common origin, i.e. due to EPI. In this article, the EPI of the 2D nanocomposite of GO-hBN has been tuned as a function of GO content (%). Tuning the EPI can effectively transform the intrinsic optical properties of a 2D material, which helps improve optoelectronic device performance. To observe the EPI tuning, a proportional parameter $k$ is evaluated as the ratio of the intensity of specific Raman mode ($I$) and $q^2$, explaining the disorders' effect on Raman line shape. This experimental work reports a linear relationship between $E_u$ and the proportional parameter ($k$) of the GO-hBN nanocomposites for the first time. We envisage that the relation $k \propto E_u$ is significant for studying the underlying physical phenomenon of EPI and modification of $E_{e-p}$ in a material. The output of our results is essential for optimizing the properties of GO-hBN nanocomposites for specific applications such as optoelectronic device design, optimizing its performance over time and for a range of electrical, optical, catalytic, and sensing applications.

**Acknowledgements**

Vidyotma Yadav is thankful to CSIR for support through fellowship. The authors thank Advanced Instrumentation Research Facility (AIRF), JNU, for SEM and Raman spectroscopic measurements. The authors also acknowledge Dr Sobhan Sen for UV-Vis spectroscopic measurements for giving access to the Spec lab at SPS, JNU. Dr Tanuja Mohanty is grateful to DST-SERB project no. CRG/2022/007270 for consumable support.

# Figure captions

**Figure 1.** Schematic of (a) energy band diagram of hBN (b) energy band diagram of GO-hBN nanocomposite and electron transfer pathways (c) Generation of Urbach tails and band gap variation due to defect introduction

**Figure 2.** Schematic diagram of the emergence of the Fano resonance effect with the representation of (a) discrete phononic state, (b) continuum of electronic states, (c) antisymmetric mix states, (d) special cases of Fano line shape for q = 0 and q = ∞.

**Figure 3.** Schematic of stepwise synthesis procedure of GO-hBN nanocomposites.

**Figure 4.** Schematic of GO-hBN nanocomposite.

**Figure 5.** SEM images of (a) hBN (b) GO-hBN (1:2) (c) GO-hBN (1:1) (d) GO-hBN (2:1), (e) GO-hBN (3:1) and (f) GO

**Figure 6.** (a) UV-Vis absorption spectra of hBN and the GO-hBN nanocomposites (b) Optical bandgap of GO-hBN nanocomposites with GO content (%) using Tauc's relation (inset: Bandgap of hBN using Tauc's plot). Indirect bandgap was attributed to hBN and GO domains in the nanocomposites. The intercept of the dash lines with the horizontal axis elucidates the bandgap value.

**Figure 7.** (a) Comparison of optical bandgap ($E_g$) and Urbach energy ($E_u$) of GO-hBN nanocomposites with GO content (%). (b) Curve fitting of bandgap in the nanocomposites with GO content (%).

**Figure 8.** Raman spectra of (a) hBN, GO and the nanocomposites (b)-(f) BWF fitting of Raman spectra of the nanocomposites as (b) GO-hBN (1:3) (c) GO-hBN (1:2) (d) GO-hBN (1:1) (e) GO-hBN (2:1) (f) GO-hBN (3:1).

**Figure 9.** (a) Comparison of $q$ and $E_u$ with GO content (%). (b) The parabolic trend of $E_{e\text{-}p}$ in terms of $E_u$ (i.e. $\frac{2E_u}{3\,K_BT}$) and the strength of EPI (i.e. $\frac{1}{|q|}$) with GO content (%).

**Figure 10.** Linear relationship of $k$ parameter with $E_u$ in GO-hBN nanocomposites.

# Table captions

**Table. 1.** Calculated values of $E_g$, $E_u$ and $q$ of pristine hBN and GO-hBN nanocomposites with different GO content (%).